\documentclass[journal=nalefd,manuscript=letter]{achemso}

\usepackage[T1]{fontenc}       
\usepackage{textcomp}
\usepackage{xcolor}
\usepackage{soul}

\author{Christoph Neumann}
\affiliation{JARA-FIT and 2nd Institute of Physics, RWTH Aachen University, 52074 Aachen, Germany}
\alsoaffiliation{Peter Gr\"unberg Institute (PGI-9), Forschungszentrum J\"ulich, 52425 J\"ulich, Germany}

\author{Leo Rizzi}
\affiliation{JARA-FIT and 2nd Institute of Physics, RWTH Aachen University, 52074 Aachen, Germany}

\author{Sven Reichardt}
\affiliation{JARA-FIT and 2nd Institute of Physics, RWTH Aachen University, 52074 Aachen, Germany}
\alsoaffiliation{Physics and Materials Science Research Unit, Universit\'e du Luxembourg, 1511 Luxembourg, Luxembourg}

\author{Bernat Terr\'es}
\affiliation{JARA-FIT and 2nd Institute of Physics, RWTH Aachen University, 52074 Aachen, Germany}
\alsoaffiliation{Peter Gr\"unberg Institute (PGI-9), Forschungszentrum J\"ulich, 52425 J\"ulich, Germany}

\author{Timofiy Khodkov}
\affiliation{JARA-FIT and 2nd Institute of Physics, RWTH Aachen University, 52074 Aachen, Germany}
\alsoaffiliation{Peter Gr\"unberg Institute (PGI-9), Forschungszentrum J\"ulich, 52425 J\"ulich, Germany}

\author{Kenji Watanabe}
\affiliation{National Institute for Materials Science,1-1 Namiki, Tsukuba, 305-0044, Japan}

\author{Takashi Taniguchi}
\affiliation{National Institute for Materials Science,1-1 Namiki, Tsukuba, 305-0044, Japan}

\author{Bernd Beschoten}
\affiliation{JARA-FIT and 2nd Institute of Physics, RWTH Aachen University, 52074 Aachen, Germany}

\author{Christoph Stampfer}
\email{stampfer@physik.rwth-aachen.de}
\affiliation{JARA-FIT and 2nd Institute of Physics, RWTH Aachen University, 52074 Aachen, Germany}
\alsoaffiliation{Peter Gr\"unberg Institute (PGI-9), Forschungszentrum J\"ulich, 52425 J\"ulich, Germany}

\title{Spatial control of laser-induced doping profiles in graphene on hexagonal boron nitride}

\abbreviations{Gr,hBN,CNP}

\begin{document}


\begin{tocentry}

%
%
%

\includegraphics{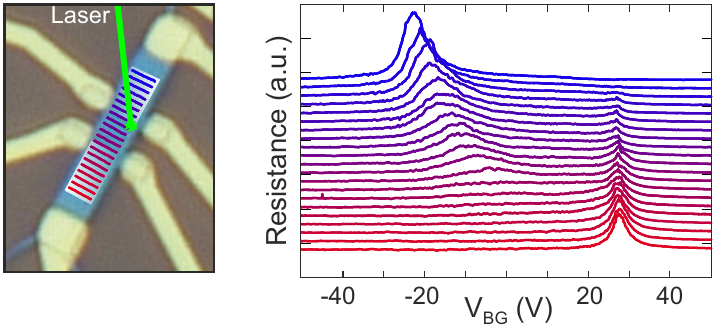}

\end{tocentry}


\begin{abstract}

We present a method to create and erase spatially resolved doping profiles in graphene-hexagonal boron nitride (hBN) heterostructures. The technique is based on photo-induced doping by a focused laser and does neither require masks nor photo resists.
This makes our technique interesting for rapid prototyping of unconventional electronic device schemes, where the spatial resolution of the rewritable, long-term stable doping profiles is only limited by the laser spot size ($\approx$ 600~nm) and the accuracy of sample positioning. Our optical doping method offers a way to implement and to test different, complex doping patterns in one and the very same graphene device, which is not achievable with conventional gating techniques.

\end{abstract}


In recent years, the stacking of two-dimensional materials bound by van~der~Waals interaction has emerged as an interesting approach for designing and studying novel device concepts for electronic \cite{ponomarenko2013,dean2013,hunt2013,woods2014} and optoelectronic applications \cite{britnell2013,withers2015,lu2014}.
In particular, material stacks built around graphene (Gr) promise interesting electronic properties \cite{decker2011,wang2013,engels2014b,engels2014,neumann2015} and many possibilities for hosting high-quality devices.
Different two-dimensional materials have been shown to be favorable substrates for graphene in such stacks \cite{kretinin2014}.
Most prominently, hexagonal boron nitride (hBN) has been used in numerous studies to encapsulate graphene and enable high-quality graphene devices \cite{dean2010,britnell2012,couto2014,engels2014,drogeler2014}.
In order to enable functional electronic devices, local gates are usually used to implement p-n junctions or other doping profiles. 
Recently, an alternative way has been reported for changing the charge carrier doping in hBN-Gr stacks by optical illumination \cite{ju2014}.
Notably this photo-induced doping in heterostructures of graphene and hBN \cite{ju2014} is significantly more efficient than for graphene on SiO$_2$ \cite{kim2013b,tiberj2013}.
Here we show that a focused laser can be used to create charge doping patterns, such as lateral p-n junctions, with high spatial precision and long lifetimes in Gr-hBN heterostructures. Importantly the presented laser-induced doping technique works completely without masks and photo-resists.
We show that the lateral resolution is essentially only limited by the laser spot size and accuracy of sample positioning.
The process maintains the high electronic mobility of the graphene sample and offers distinct advantages over conventional gate electrodes, such as rewritability, the possibility of changing and controlling doping profiles in a single device, and a reduction of process steps.
This makes our findings highly interesting for prototyping unconventional electronic device schemes based on graphene.

Our sample consists of a single-layer graphene sheet which is encapsulated by two multi-layer hBN flakes.
This material stack is obtained by a dry and resist-free transfer process, which has been shown to result in high-mobility graphene devices \cite{wang2013,engels2014b,banszerus2015}.
The resulting heterostructure is placed on a highly doped silicon substrate with a 285~nm thick SiO$_{2}$ layer (Figure 1a).
The sample is structured into a Hall bar using electron beam lithography and reactive ion etching.
Finally, it is contacted via chrome/gold electrodes on the sides of the heterostructure \cite{wang2013}. The width of the Hall bar is 1.9~\textmu m and the total length is 12.3~\textmu m.
For transport measurements, we use a four-probe geometry with a constant source-drain current of 50~nA (Figure 1b).
The longitudinal voltage $V_{\mathrm{xx}}$ is probed at the two lower contacts.
Our optical setup consists of a confocal laser system with a $100\times$ objective.
For excitation, we use a laser with energy $E_{\mathrm{L}}=2.33$~eV and an intensity between 2 and 4~mW.
Reflected and scattered light is detected via a single mode optical fiber and a spectrometer with a 1200~lines/mm grating.
The setup enables us to locally investigate the Raman signal of the sample at 4.2~K.
A typical Raman spectrum obtained on the Hall bar is presented in Figure~1c.
The characteristic graphene Raman G~line (around 1580~cm$^{-1}$) and the 2D~line (around 2680~cm$^{-1}$) are observed as well as a line at around 1365~cm$^{-1}$, which originates from E$_{\mathrm{2g}}$ phonons in the hBN layers.
In particular, the small full width at half maximum (FWHM) of the graphene 2D~line of around 17~cm$^{-1}$ is an indication of the high crystal quality and local flatness of the encapsulated graphene sheet \cite{neumann2015,neumann2015b}.

\begin{figure}
\centering
\includegraphics[draft=false,keepaspectratio=true,clip, width=0.5\linewidth]{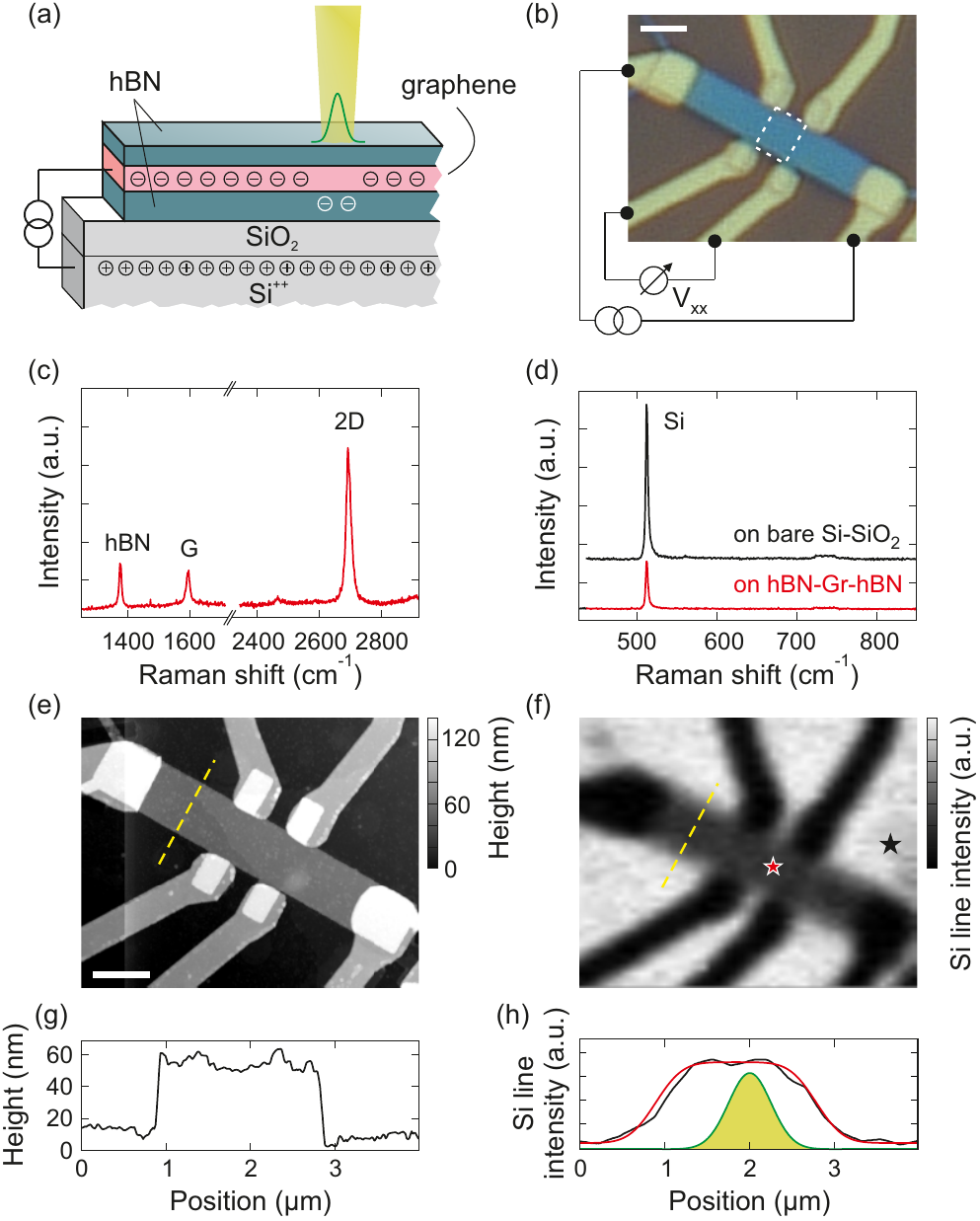}
\caption[FIG1]{
(a) Schematic illustration of an hBN-Gr-hBN sandwich on a Si$^{++}$/SiO$_2$ substrate. The sample can be locally illuminated by a laser ($E_{\mathrm{L}}=2.33$~eV). In the illuminated area, the hBN is charged screening away the electric field of the back gate.
(b) Optical image of an hBN-Gr-hBN Hall bar device. The longitudinal voltage $V_{\mathrm{xx}}$ is probed between the two bottom contacts. The scale bar represents 2~\textmu m. The white, dashed rectangle marks the area where the voltage is probed.
(c) Raman spectrum taken on the Hall bar at the position of the red star in panel (f). The prominent G and 2D~peaks of Gr are seen at around 1580~cm$^{-1}$ and 2680~cm$^{-1}$, respectively.
The Raman line at 1365~cm$^{-1}$ comes from the hBN substrate.
(d) Raman spectra obtained on the uncovered substrate next to the Hall bar (black star in panel (f)) and on the Hall bar (red star in panel (f)). The characteristic silicon peaks have a higher amplitude on the uncovered wafer as compared to the hBN-Gr-hBN covered area.
(e) SFM image of the Hall bar device. The scale bar represents 2 \textmu m.
(f) Scanning Raman microscopy image of the device. The intensity of the silicon peak at 520~cm$^{-1}$ is color encoded.
(g) Line cut of the SFM image along the yellow, dashed line in panel (e).
(h) Line cut of the scanning Raman microscopy image along the yellow, dashed line in panel (f) (black line). The red line represents the convolution of a step function with a Gaussian profile with a standard deviation of 250~nm (green curve).
}
\label{fig1}
\end{figure}

A scanning force microscopy (SFM) image of the Hall bar is shown in Figure~1e.
Comparison with a scanning Raman microscopy image (Figure~1f), showing the intensity of the prominent Si Raman line at 520~cm$^{-1}$ (see spectra in Figure~1d), reveals the high spatial resolution of our optical setup.
The silicon peak intensity is reduced when the substrate is covered by the hBN-Gr heterostructure (compare the red and black curves in Figure 1d) and is completely suppressed in metal-coated areas (i.e. on the gold electrodes).
Imaging the Si peak intensity allows us to precisely navigate across the device, which is needed for the subsequent experiments.
By comparing line cuts of the SFM and scanning Raman microscopy images across the Hall bar (compare Figures~1g and 1h), we can further gain a quantitative measure for the laser spot size.
Assuming a Gaussian laser profile and convoluting it with a step function, we adjust the width of the Gaussian so that the convoluted profile matches the Raman line cut (see Figure~1h).
This way, we estimate the standard deviation of the Gaussian to be around 250~nm (green curve in Figure 1h), which corresponds to a FWHM of $\sim 590$~nm.
This value sets the size of our laser spot and limits the spatial resolution of the doping profiles discussed below.

\begin{figure}
\centering
\includegraphics[draft=false,keepaspectratio=true,clip,width=0.8\linewidth]{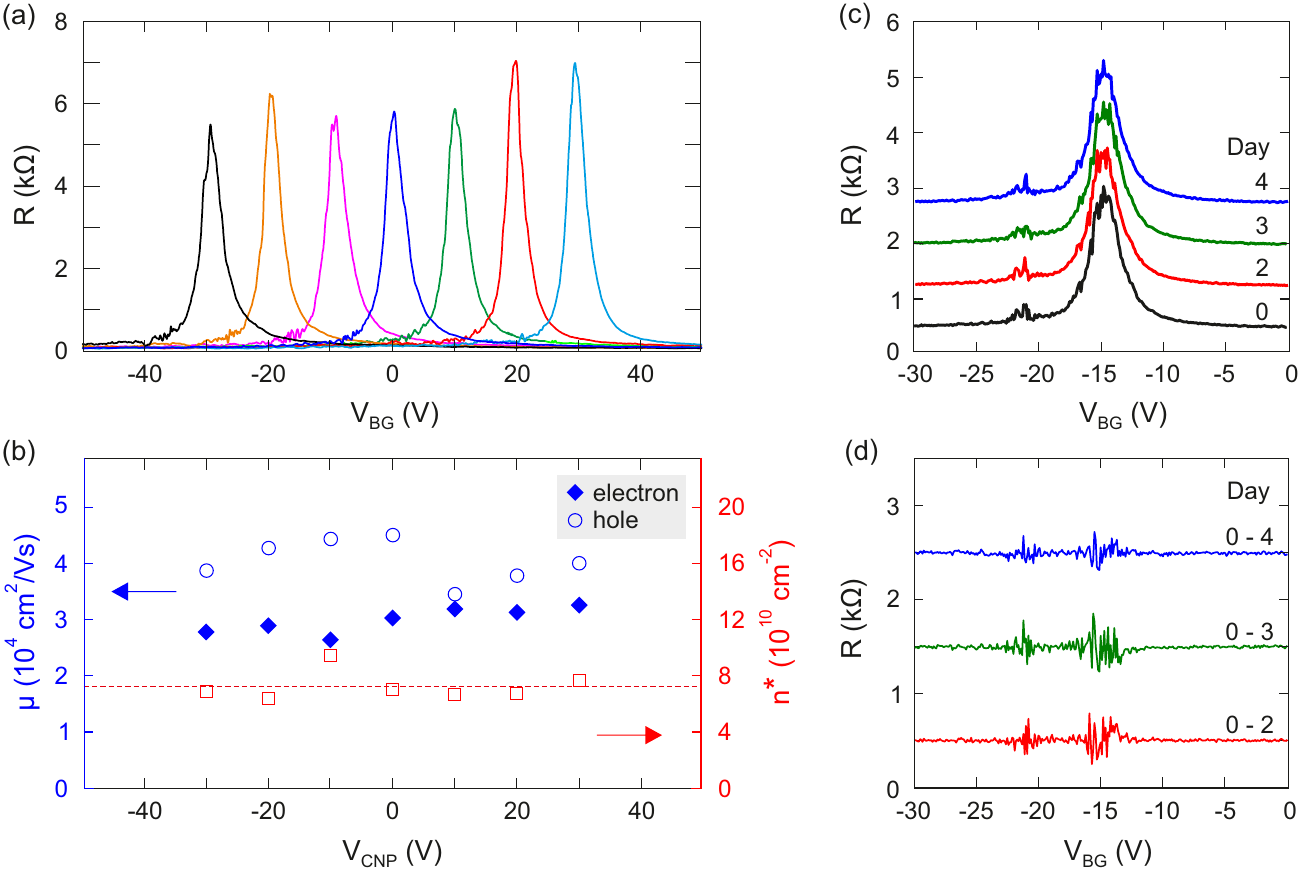}
\caption[FIG2]{
(a) Transport measurements highlighting the photo-induced doping effect. Seven back gate characteristics of the device with different CNPs ranging from -30 to 30~V.
(b) Electron and hole mobilities ($\mu_{\mathrm{e}}$ and $\mu_{\mathrm{h}}$) and charge carrier density fluctuations around the CNP ($n^*$) as extracted from the curves in panel (a). All three quantities remain mostly constant under application of the photo-induced doping effect.
(c) Back gate characteristics of the device taken immediately after a doping profile was written and 2, 3, and 4 days later. The traces  are offset for clarity by 750 $\Omega$.
(d) The traces from panel c recorded after 2, 3, and 4 days are subtracted from the curve measured immediately after the doping profile was written. The traces are offset by 1~k$\Omega$ each.
}
\label{fig2}
\end{figure}

The combination of an electrically contacted device with a confocal laser setup allows us to investigate the spatial control of the laser-induced doping profiles in Gr-hBN heterostructures.
Ju and coworkers \cite{ju2014} attributed the photo-induced doping effect in hBN-Gr structures to nitrogen vacancies and carbon defects in the hBN, which give rise to defect states deep in the band gap with energies of 2.8~eV (nitrogen vacancy) and 2.6~eV (carbon impurity) \cite{attaccalite2011}.
They argue that with the help of photons, these states can be occupied by charge carriers injected from the gated graphene\cite{ju2014}.
When the laser is turned on, charge carriers occupying defect states in the hBN layer are excited and, due to the applied gate voltage, move toward the graphene sheet, leaving behind oppositely charged states in the hBN layer.
This process continues until the back gate is fully screened by the increasingly charged hBN layer (see also Figure~1a).
When turning the laser off again, this effect has effectively shifted the charge neutrality point (CNP) of the graphene sheet to the chosen value of the back gate voltage. Our laser energy of 2.33~eV in combination with the rather high laser intensity (2 to 4~mW) leads to high doping rates, such that only short illumination times are required to shift the CNP.

We expect that the strong photo-induced doping effect in hBN-Gr stacks on SiO$_2$ is closely related to the asymmetric gate oxide structure, i.e. the presence of the insulating SiO$_{2}$ layer underneath the hBN. The charges injected in the hBN likely diffuse to the bottom of the hBN layer in the direction of the electric field from the Si$^{++}$ back gate until they are stopped at the SiO$_2$/hBN interface. However, further investigations on the physical mechanisms of the photo-induced doping effect require different sample geometries and spectroscopy analysis techniques, which are beyond the scope of this manuscript. In this work, we focus on the creation, the erasing, the rewriting, and most importantly the spatial resolution of laser-induced doping profiles in this material stack, which do not rely on the use of photo-resists or masks.

In Figure~2a we show that this effect can be employed to shift the CNP to arbitrary negative as well as positive back gate voltages.
To set the CNP of the entire graphene Hall bar to a specific value, we apply the corresponding back gate voltage and scan over the full area depicted in Figure~1f with a total illumination time of 120~s.
The typical back gate characteristic of graphene \cite{novoselov2004} is maintained in all cases.
From each back gate characteristic, we extract the electron mobility $\mu_{\mathrm{e}}$, the hole mobility $\mu_{\mathrm{h}}$, and the charge carrier density fluctuation $n^*$.
The latter has been extracted by the method described in Reference~\citenum{couto2014} and is a good measure for the electronic disorder (i.e. electron-hole puddles) in bulk graphene.
We find average values of $\mu_{\mathrm{e}} = 30,000$~cm$^2$/(Vs), $\mu_{\mathrm{h}} = 40,000$~cm$^2$/(Vs), and $n^* = 7 \times 10^{10}$~cm$^{-2}$ (see Figure 2b).
Importantly, $\mu_{\mathrm{e}}$, $\mu_{\mathrm{h}}$, and $n^*$ remain mostly constant for each chosen CNP, which shows the non-destructive nature of this process (Figure~2b).
After turning the laser off, the back gate characteristic at 4.2~K remains unchanged for at least four days, which is the longest we have waited without manipulating the doping profile of the sample (see Figures~2c and d).

\begin{figure}
\centering
\includegraphics[draft=false,keepaspectratio=true,clip,width=1\linewidth]{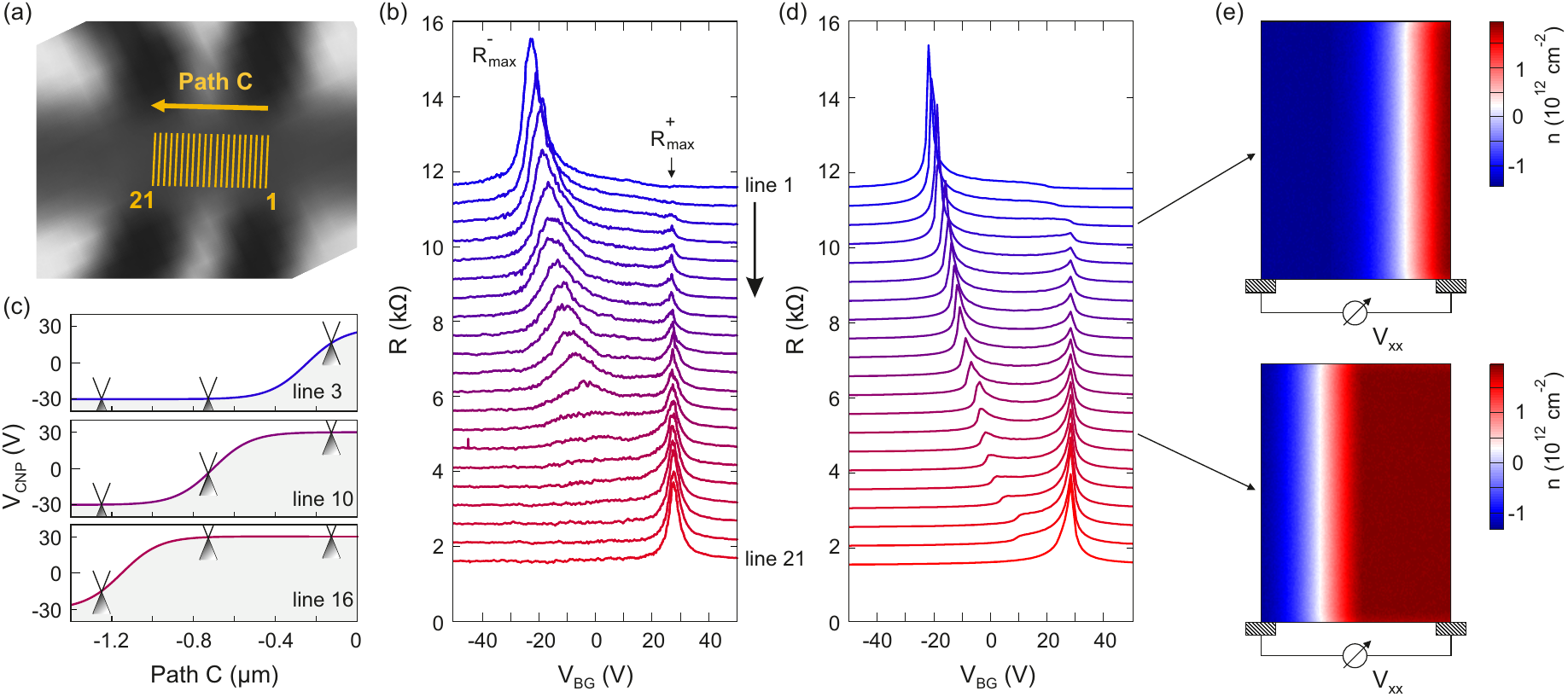}
\caption[FIG3]{
(a) Scanning, confocal Raman image of the integrated Si~peak intensity. The orange lines indicate the parts written by the laser at a back gate voltage of +30~V to tune the area step by step from an n-doped regime via a local p-n junction to an entirely p-doped regime.
(b) Back gate characteristics of the Hall bar measured after each individual line written with the laser (first back gate trace is at the top). The lines are offset for clarity by 500 $\Omega$ each (starting from line 21).
(c) Illustration of the value of $V_{\mathrm{CNP}}$ along the horizontal extent of the sample for three different doping configurations. From top to bottom, the sample is increasingly more p-doped.
(d) Simulated back gate traces. The parameters used are $\mu_{\mathrm{e}} = 30,000$~cm$^2$/(Vs), $\mu_{\mathrm{h}} = 40,000$~cm$^2$/(Vs), and $n^* = 7 \times 10^{10}$~cm$^{-2}$. The standard deviation of the Gaussian used for the doping profile is 300~nm. From top to bottom a greater part of the four-probe contact area is set to +30~V. The lines are offset for clarity.
(e) Two exemplary doping profiles used in the simulation.
}
\label{fig3}
\end{figure}

\begin{figure*}
\centering
\includegraphics[draft=false,keepaspectratio=true,clip,width=0.5\linewidth]{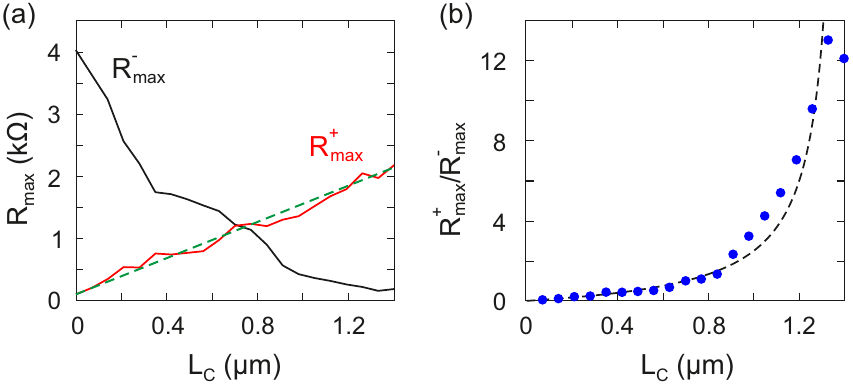}
\caption[FIG4]{
(a) The maximum resistance for the right ($R^+_{\mathrm{max}}$, red curve) and left ($R^-_{\mathrm{max}}$, black curve) resistance peak of each trace shown in Figure 3b are extracted and plotted against the absolute value of the position of the last line written by the laser along ''Path C" $L_{C}$ indicated in Figure 3a. The green, dashed line is a linear fit with a slope of 1.4~$\Omega$/nm highlighting the expected linear trend of $R^+_{\mathrm{max}}$ considering simple, diffusive transport.
(b) The ratio of the peak resistances shown in panel (a) ($R^+_{\mathrm{max}}$/$R^-_{\mathrm{max}}$) is plotted against $L_{C}$ (blue points). The dashed, black line is given by the expression $L_{C}/(L_{0} - L_{C})$ with $L_0 = 1.4$~\textmu m. It shows the behavior from assuming that the graphene sheet consists of two resistances in series, one for the n-doped and one for the p-doped regime.
}
\label{fig4}
\end{figure*}

After demonstrating that we can shift the CNP of the entire Hall bar, we next focus on defining spatially varying doping patterns within the graphene sheet.
This aspect is highly interesting for potential applications as it enables the definition of re-writable, in-plane p-n and n-p-n junctions with any desired pattern, size, and orientation.
To examine the spatial precision of our approach, we start by continuously moving the CNP in the sensitive four-probe area from -30~V to +30~V.
In the beginning, the CNP of the entire sample is set to -30~V by scanning over the sample with our laser.
Simultaneously, we record a Raman image of the prominent Si line.
The integrated Si~line intensity is displayed in Figure~3a.
In the following, the laser is turned off and the back gate voltage is set to +30~V.
Afterward, the laser is turned on for writing a single line across the Hall bar on the very right side of the four-probe area (see vertical line 1 in Figure~3a).
The total illumination time during the writing of the line was 7~s.
After switching the laser off again, the back gate characteristic of the device is probed, resulting in the upper curve in Figure~3b.
This procedure is repeated 20 times.
In every iteration, the written line is intentionally shifted by $\sim$70~nm to the left (vertical arrow and ``Path C'' in Figure~3a).
The resistance curves obtained after each iteration are displayed in Figure~3b, with the lines offset for clarity.
With each line, the CNP of a greater part of the sensitive area is set to +30~V, resulting in the appearance of a typical graphene p-n junction back gate characteristic \cite{williams2007} (middle part of Figure~3b).
Additionally, stray light from the laser leads to a continuous doping of the Hall bar and as a result the initial Dirac peak is gradually broadened and shifted toward positive voltages.
This effect is especially strong after writing the first line, when the initial Dirac peak is not anymore at -30~V but is shifted to -23~V.
Finally, after the entire four-probe area was covered with lines written by the laser, the initial, left Dirac peak has completely vanished and only the new CNP at +30~V can be seen in the back gate characteristic (line 21 in Figure~3b).
The steady and continuous shift from an entirely electron-doped four-probe area over a p-n junction to an entirely hole-doped area (see evolution from top to bottom panel in Figure~3c) demonstrates the high spatial precision with which doping profiles can be made with this technique.

To cross-check the measurements, we employ a simple, purely diffusive charge transport model, in which we divide the area where $V_{\mathrm{xx}}$ is probed (1.4 \textmu m $\times$ 1.9 \textmu m; compare white, dashed rectangle in Figure~1b) into individual squares with a size of 10~nm $\times$ 10~nm each.
To each square we then assign a charge distribution

\begin{equation}
n_{ij}(V_{\mathrm{BG}}) = n_{\mathrm{dop},ij} + n_{\mathrm{gate}}(V_{\mathrm{BG}}) + n^*_{ij},
\label{eqn:doping}
\end{equation}

consisting of three contributions.
The first contribution, $n_{\mathrm{dop},ij}$, is the spatially varying doping profile due to the photo-induced doping effect.
Each written laser line is modeled as a Gaussian with a standard deviation of 300~nm, which slightly differs from the experimentally determined value (250~nm, see Figure~1h) to account for various experimental uncertainties as described below.
Additionally, we add a constant offset to incorporate the initial homogeneous photo-induced doping of the sample.
We adjust this constant offset for each trace to account for the increasing stray light-induced doping of the sample.
The second contribution, $n_{\mathrm{gate}}(V_{\mathrm{BG}}) = \alpha V_{\mathrm{BG}}$, accounts for the field-effect-induced charge carrier density during a back gate sweep, with $\alpha = 5.5 \times 10^{10}$~V$^{-1}$cm$^{-2}$ being the capacitive coupling constant of the back gate as extracted from quantum Hall measurements (not shown).
The final contribution, $n^*_{ij}$, represents the built-in charge density variations across the sample.
In our model, we implement it by assigning a randomly generated number uniformly distributed between $-n^*/2$ and $n^*/2$ to each square.
Finally, we compute the conductivity of every square by

\begin{equation}
\sigma_{ij} = e \mu |n_{ij}| + \sigma_0,
\label{eqn:conductivity}
\end{equation}

where $e$ is the elementary charge, $n_{ij}$ the charge carrier density of the square and $\sigma_0$ is a residual conductivity to account for the finite resistance of graphene at the charge neutrality point adjusted to $10^{-4}$~S to reproduce the data.
Furthermore, $\mu$ is the electron/hole mobility, depending on the sign of $n_{ij}$, for which we use the average values of $\mu_{\mathrm{e}} = 30,000$~cm$^{-2}$/(Vs) and $\mu_{\mathrm{h}} = 40,000$~cm$^{-2}$/(Vs), as obtained from the six back gate traces in Figure~2a.
By applying Ohm's law to each square and relating the currents and voltages of each square via Kirchhoff's laws, we calculate the total resistance of the graphene sheet from the applied bias voltage and the total outgoing current.

We end up with the simulated back gate characteristics shown in Figure~3d, which match the experimental results quite well.
In Figure~3e the corresponding doping profiles of $n_{\mathrm{dop},ij}$ across the four-probe area for two exemplary back gate traces are visualized.
The fact that our simple model is able to reproduce the most prominent features of our measurements clearly indicates that the photo-induced doping effect can indeed be effectively used to define spatially resolved doping patterns.

In Figure 4a, we show the maximum resistance value measured at around +30~V back gate voltage $R^+_{\mathrm{max}}$ (red trace) as well as the maximum value obtained for negative back gate voltage $R^-_{\mathrm{max}}$ (black trace) as highlighted in Figure~3b. Both peak resistances are plotted against $L_{C}$ , which is defined as the absolute value of the position of the last, exposed line along ''Path C" prior to measuring the corresponding back gate trace. We observe that with increasing $L_{C}$, $R^+_{\mathrm{max}}$ increases linearly, while $R^-_{\mathrm{max}}$ drops. In particular, the linear increase of $R^+_{\mathrm{max}}$ is in good agreement with simple, diffusive transport considerations, where $R^+_{\mathrm{max}}$ is completely dominated by the length of the sample area with a Dirac point at +30~V. This area increases linearly with $L_C$, explaining the dependence of $R^{+}_{max} \propto L_{C}$. From a linear fit to the data (see green, dashed line in Figure 4a) we extract a resistance change per length of $\Delta R^{+}_{max}/\Delta L_{C} = 1.4$~$\Omega$/nm. 
In Figure 4b the ratio of both resistance peaks $R^{+}_{max}/R^{-}_{max}$ is shown (blue points). The continuous increase can be understood from simple diffusive transport considerations. In this case, the ratio of the resistance peaks is given by $R^{+}_{max}/R^{-}_{max}=L_{C}/(L_{0}- L_{C})$, where $L_{0}= 1.4$~\textmu m. This value highlights the very high sensitivity in resistance change as function of exposed area (i.e. length), which might be of interest for future applications. This expression describes the resistance ratio without any free parameters. The corresponding curve is shown in Figure 4b (dashed, black line). It shows excellent agreement with our measurements. From this representation it is evident that we continuously change the doping profile along ''Path C" upon writing individual lines with a step width of $\sim$70~nm and very high spatial precision.
Our results show that the resolution of this technique is essentially only limited by the size of the laser spot and the accuracy of sample positioning.

\begin{figure*}
\centering
\includegraphics[draft=false,keepaspectratio=true,clip,width=1\linewidth]{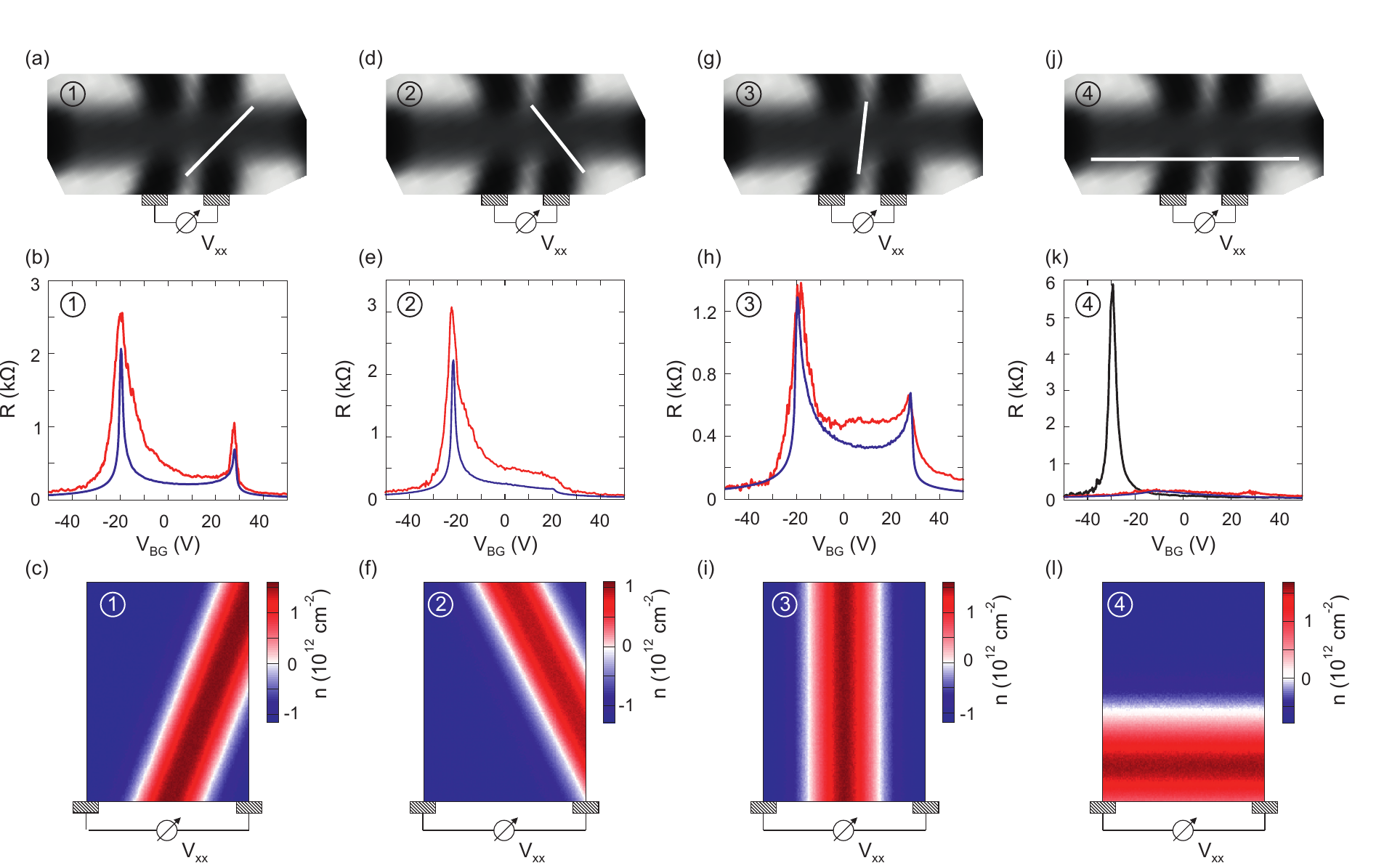}
\caption[FIG5]{
(a) Scanning Raman image of the integrated Si~peak intensity. The line indicates the doping pattern that was written with the laser.
(b) The measured (red) and simulated (blue) back gate characteristics after writing line~1 (see panel (a)). The initial CNP, which was set to -30~V, is shifted to more positive values due to stray light from the laser when writing the line at +30~V.
(c) Doping profile of line~1 (compare panel (a)) as used in the model trace shown in panel (b). The standard deviation of the Gaussian doping profile was set to 200~nm.
(d,e,f) The same as panels (a,b,c) where line~2 (see panel (d)) was written with the laser after resetting the entire device to a CNP of -30~V (see main text). The standard deviation of the Gaussian doping profile (see panel (f)) used in the model was also set to 200~nm.
(g,h,i) The same as panels (a,b,c) where line~3 (see panel (g)) was written with the laser after resetting the entire device to a CNP of -30~V (see main text). The standard deviation of the Gaussian doping profile (see panel (i)) used in the model was set here to 230~nm.
(j,k,l) The same as panels (a,b,c) where line~4 (see panel (j)) was written with the laser after resetting the entire device to a CNP of -30~V (see main text). The black line in panel (k) represents the measured back gate characteristic prior to writing line~4, when the entire sample was still set to $V_{\mathrm{CNP}}=-30$~V. In this case, the standard deviation of the Gaussian doping profile (see panel (l)) used in the model was set to 330~nm.
}
\label{fig5}
\end{figure*}

The high spatial precision of the laser-induced doping technique allows the writing and erasing of distinct, well-defined doping profiles.
We demonstrate this by writing four different lines with the laser as illustrated in Figures~5a, 5d, 5g, and 5j.
In all four cases, the entire sample was first set to a CNP of -30~V before the individual lines were written at a back gate voltage of +30~V.
For each case, we cross-check the measurements by using the simple diffusive model described above (and used in Figure 3d) to simulate a similar doping geometry.
As before, we adapt the constant doping offset in the simulation to account for stray light from the written lines at +30~V that slightly shifts the -30~V CNP in the rest of the device.
We also adjust the width of the simulated Gaussian profile to compensate for uncertainties during the measurement such as a possible misalignment of the optical path and positioning uncertainties due to creeping effects of the piezo stage.
Adjusting the width of the Gaussian also partly compensates for the fact that the actual shape of the doping profile induced by a single laser line is not exactly a Gaussian.
Once enough charge carriers in the hBN have been excited so that the hBN layer completely screens away the back gate-induced electric field, the photo-doping effect stops and hence the actual doping profile most likely has a saturation area in the center, where the intensity is highest.
A good Gaussian profile might only be obtained by perfectly adjusting illumination time and laser intensity to one another.

We start by writing line~1 (see Figure~5a), introducing a p-n junction between the two contacts where $V_{\mathrm{xx}}$ is probed.
The resulting back gate characteristic is shown in red in Figure~5b, showing the typical p-n junction behavior with two resistance peaks at -20~V and +30~V.
It is again evident that stray light plays an important role, as the initial, left CNP is shifted from the original -30~V to -20~V and is also more smeared out than the newly written CNP at positive gate voltage.
To confirm the p-n junction doping profile, we calculate the back gate characteristic of a comparable doping pattern (shown in Figure~5c).
The resulting trace (blue curve in Figure~5b) reproduces the main features of the measured trace quite well.
After erasing the written doping profile and resetting the CNP of the entire Hall bar to -30~V, line~2 is written, again at a back gate voltage of +30~V.
As seen in Figure~5d, there is now a conducting path connecting the relevant contacts that is not crossed by the laser line and consequently has its expected CNP at the original -30~V.
The measured trace (red line in Figure~5e) indeed shows only one distinct peak at negative gate voltage.
This is further backed up by the simulated trace (blue trace in Figure~5e) using the doping profile depicted in Figure~5f.
After resetting the CNP of the Hall bar to -30~V, line~3 is written.
The measured back gate trace (see red line in Figure~5h) shows two peaks at -20~V and +30~V.
Compared to line~1, the peak resistance is significantly lower and the right peak is less distinct, which can be attributed to the above mentioned experimental uncertainties.
This is further supported by the calculation (see blue line in Figure~5h), for which the width of the Gaussian profile had to be slightly increased (see Figure~5i).
However, from the measurements it is evident that an n-p-n junction can be accurately written with this technique on a length as low as 1.4~\textmu m.
Finally, we investigate a fourth interesting doping profile.
Again, we first set the entire Hall bar to a CNP of -30~V.
The resulting back gate characteristic with a single Dirac peak at -30~V can be seen in Figure~5k (black trace).
Afterward, line~4 is written with the laser at +30~V.
Since the laser line runs in parallel to the path along which $V_{\mathrm{xx}}$ is probed, there is now a highly doped path from source to drain for every value of the back gate voltage, which is reflected in the measured back gate trace, shown in Figure~5k (red curve).
As expected, the resistance over the whole range of the back gate voltage is much lower than the resistance measured at the CNP in the homogeneously doped case (black trace).
This is further supported by the calculated back gate trace (blue line in Figure~5k) using the doping profile shown in Figure~5l.
In all four cases, the qualitative match between the measured and simulated traces indicates that the written laser lines indeed produce the desired doping geometries.

In conclusion, we demonstrated that controllable, laser-induced doping in graphene-hBN heterostructures can be used to define long-term stable doping patterns such as p-n and n-p-n junctions with high spatial resolution and without the use of masks and photo-resists.
The CNP of the device can be shifted to any desired value of the back gate voltage without reducing the charge carrier mobility.
We focused on the spatial resolution of this technique and demonstrated the continuous evolution of an entirely n-doped graphene regime via a local p-n junction to a fully p-doped regime by step-by-step illumination of small areas of the graphene Hall bar with a green laser.
Cross-checking the data with a simple, fully diffusive transport model further underlines that our technique is in principle only limited by the laser spot size (Gaussian standard deviation of 250~nm) and the accuracy of the piezo stage, making it comparable to conventional optical lithography techniques.
Consequently, we wrote and erased different doping pattern geometries in a single Hall bar device, creating e.g. a custom-made n-p-n junction.
This optical approach to locally dope high-quality graphene devices has distinct advantages over conventional techniques that make use of local gate electrodes in terms of rewritability and a reduction of the amount of difficult and possibly invasive lithography, etching, and contacting steps \cite{wang2013,engels2014b}.
Additionally, doping patterns with numerous p-n junctions could be implemented and erased in a single device, opening unprecedented opportunities to study graphene-based electron optic devices \cite{rickhaus2013,young2009,taychatanapat2015}, Veselago lenses \cite{cheianov2007}, and Klein tunneling phenomena\cite{beenakker2008}.
Moreover, other complex doping profiles which are difficult or impossible to realize with conventional gate electrodes, like isolated, localized doping spots and arrays, can be realized with this technique.
Likely, the photo-induced doping effect can be extended to other van~der~Waals heterostructures based on hBN, opening a wide range of possible studies on the creation of lateral transition-metal-dichalcogenide-hBN transistors without local gate electrodes.


\begin{acknowledgement}

The authors thank M. Dr\"ogeler and F. Hassler for helpful discussions.
Support by the Helmholtz Nanoelectronic Facility (HNF), the Deutsche Forschungsgemeinschaft, the ERC (GA-Nr. 280140), and the EU project Graphene Flagship (contract no. NECT-ICT-604391), are gratefully acknowledged.
S. R. acknowledges funding by the National Research Fund (FNR) Luxembourg.
\end{acknowledgement}

%
%


\providecommand*\mcitethebibliography{\thebibliography}
\csname @ifundefined\endcsname{endmcitethebibliography}
  {\let\endmcitethebibliography\endthebibliography}{}


\begin{mcitethebibliography}{31}
\providecommand*\natexlab[1]{#1}
\providecommand*\mciteSetBstSublistMode[1]{}
\providecommand*\mciteSetBstMaxWidthForm[2]{}
\providecommand*\mciteBstWouldAddEndPuncttrue
  {\def\EndOfBibitem{\unskip.}}
\providecommand*\mciteBstWouldAddEndPunctfalse
  {\let\EndOfBibitem\relax}
\providecommand*\mciteSetBstMidEndSepPunct[3]{}
\providecommand*\mciteSetBstSublistLabelBeginEnd[3]{}
\providecommand*\EndOfBibitem{}
\mciteSetBstSublistMode{f}
\mciteSetBstMaxWidthForm{subitem}{(\alph{mcitesubitemcount})}
\mciteSetBstSublistLabelBeginEnd
  {\mcitemaxwidthsubitemform\space}
  {\relax}
  {\relax}

\bibitem[Ponomarenko et~al.(2013)Ponomarenko, Gorbachev, Yu, Elias, Jalil,
  Patel, Mishchenko, Mayorov, Woods, Wallbank, M, Piot, Potemski, Grigorieva,
  Novoselov, Guinea, Fal{'}ko, and Geim]{ponomarenko2013}
Ponomarenko,~L. et~al.  \emph{Nature} \textbf{2013}, \emph{497}, 594--597\relax
\mciteBstWouldAddEndPuncttrue
\mciteSetBstMidEndSepPunct{\mcitedefaultmidpunct}
{\mcitedefaultendpunct}{\mcitedefaultseppunct}\relax
\EndOfBibitem
\bibitem[Dean et~al.(2013)Dean, Wang, Maher, Forsythe, Ghahari, Gao, Katoch,
  Ishigami, Moon, Koshino, Taniguchi, Watanabe, Shepard, Hone, and
  Kim]{dean2013}
Dean,~C.; Wang,~L.; Maher,~P.; Forsythe,~C.; Ghahari,~F.; Gao,~Y.; Katoch,~J.;
  Ishigami,~M.; Moon,~P.; Koshino,~M.; Taniguchi,~T.; Watanabe,~K.;
  Shepard,~K.; Hone,~J.; Kim,~P. \emph{Nature} \textbf{2013}, \emph{497},
  598--602\relax
\mciteBstWouldAddEndPuncttrue
\mciteSetBstMidEndSepPunct{\mcitedefaultmidpunct}
{\mcitedefaultendpunct}{\mcitedefaultseppunct}\relax
\EndOfBibitem
\bibitem[Hunt et~al.(2013)Hunt, Sanchez-Yamagishi, Young, Yankowitz, LeRoy,
  Watanabe, Taniguchi, Moon, Koshino, Jarillo-Herrero, and Ashoori]{hunt2013}
Hunt,~B.; Sanchez-Yamagishi,~J.; Young,~A.; Yankowitz,~M.; LeRoy,~B.~J.;
  Watanabe,~K.; Taniguchi,~T.; Moon,~P.; Koshino,~M.; Jarillo-Herrero,~P.;
  Ashoori,~R. \emph{Science} \textbf{2013}, \emph{340}, 1427--1430\relax
\mciteBstWouldAddEndPuncttrue
\mciteSetBstMidEndSepPunct{\mcitedefaultmidpunct}
{\mcitedefaultendpunct}{\mcitedefaultseppunct}\relax
\EndOfBibitem
\bibitem[Woods et~al.(2014)Woods, Britnell, Eckmann, Yu, Gorbachev, Kretinin,
  Park, Ponomarenko, Katsnelson, Gornostyrev, Watanabe, Taniguchi, Casiraghi,
  Gao, Geim, and Novoselov]{woods2014}
Woods,~C. et~al.  \emph{Nature Physics} \textbf{2014}, \emph{10},
  451--456\relax
\mciteBstWouldAddEndPuncttrue
\mciteSetBstMidEndSepPunct{\mcitedefaultmidpunct}
{\mcitedefaultendpunct}{\mcitedefaultseppunct}\relax
\EndOfBibitem
\bibitem[Britnell et~al.(2013)Britnell, Ribeiro, Eckmann, Jalil, Belle,
  Mishchenko, Kim, Gorbachev, Georgiou, Morozov, Grigorenko, Geim, Casiraghi,
  Castro~Neto, and Novoselov]{britnell2013}
Britnell,~L.; Ribeiro,~R.; Eckmann,~A.; Jalil,~R.; Belle,~B.; Mishchenko,~A.;
  Kim,~Y.-J.; Gorbachev,~R.; Georgiou,~T.; Morozov,~S.; Grigorenko,~A.;
  Geim,~A.; Casiraghi,~C.; Castro~Neto,~A.; Novoselov,~K. \emph{Science}
  \textbf{2013}, \emph{340}, 1311--1314\relax
\mciteBstWouldAddEndPuncttrue
\mciteSetBstMidEndSepPunct{\mcitedefaultmidpunct}
{\mcitedefaultendpunct}{\mcitedefaultseppunct}\relax
\EndOfBibitem
\bibitem[Withers et~al.(2015)Withers, Del Pozo-Zamudio, Mishchenko, Rooney,
  Gholinia, Watanabe, Taniguchi, Haigh, Geim, Tartakovskii, and
  Novoselov]{withers2015}
Withers,~F.; Del Pozo-Zamudio,~O.; Mishchenko,~A.; Rooney,~A.; Gholinia,~A.;
  Watanabe,~K.; Taniguchi,~T.; Haigh,~S.; Geim,~A.; Tartakovskii,~A.;
  Novoselov,~K. \emph{Nature Materials} \textbf{2015}, \emph{14},
  301--306\relax
\mciteBstWouldAddEndPuncttrue
\mciteSetBstMidEndSepPunct{\mcitedefaultmidpunct}
{\mcitedefaultendpunct}{\mcitedefaultseppunct}\relax
\EndOfBibitem
\bibitem[Lu et~al.(2014)Lu, Li, Watanabe, Taniguchi, and Andrei]{lu2014}
Lu,~C.-P.; Li,~G.; Watanabe,~K.; Taniguchi,~T.; Andrei,~E.~Y. \emph{Physical
  Review Letters} \textbf{2014}, \emph{113}, 156804\relax
\mciteBstWouldAddEndPuncttrue
\mciteSetBstMidEndSepPunct{\mcitedefaultmidpunct}
{\mcitedefaultendpunct}{\mcitedefaultseppunct}\relax
\EndOfBibitem
\bibitem[Decker et~al.(2011)Decker, Wang, Brar, Regan, Tsai, Wu, Gannett,
  Zettl, and Crommie]{decker2011}
Decker,~R.; Wang,~Y.; Brar,~V.~W.; Regan,~W.; Tsai,~H.-Z.; Wu,~Q.; Gannett,~W.;
  Zettl,~A.; Crommie,~M.~F. \emph{Nano Letters} \textbf{2011}, \emph{11},
  2291--2295\relax
\mciteBstWouldAddEndPuncttrue
\mciteSetBstMidEndSepPunct{\mcitedefaultmidpunct}
{\mcitedefaultendpunct}{\mcitedefaultseppunct}\relax
\EndOfBibitem
\bibitem[Wang et~al.(2013)Wang, Meric, Huang, Gao, Gao, Tran, Taniguchi,
  Watanabe, Campos, Muller, Guo, Kim, Hone, Shepard, and Dean]{wang2013}
Wang,~L.; Meric,~I.; Huang,~P.; Gao,~Q.; Gao,~Y.; Tran,~H.; Taniguchi,~T.;
  Watanabe,~K.; Campos,~L.; Muller,~D.; Guo,~J.; Kim,~P.; Hone,~J.;
  Shepard,~K.; Dean,~C. \emph{Science} \textbf{2013}, \emph{342},
  614--617\relax
\mciteBstWouldAddEndPuncttrue
\mciteSetBstMidEndSepPunct{\mcitedefaultmidpunct}
{\mcitedefaultendpunct}{\mcitedefaultseppunct}\relax
\EndOfBibitem
\bibitem[Engels et~al.(2014)Engels, Terr{\'e}s, Klein, Reichardt, Goldsche,
  Kuhlen, Watanabe, Taniguchi, and Stampfer]{engels2014b}
Engels,~S.; Terr{\'e}s,~B.; Klein,~F.; Reichardt,~S.; Goldsche,~M.; Kuhlen,~S.;
  Watanabe,~K.; Taniguchi,~T.; Stampfer,~C. \emph{Physica Status Solidi (b)}
  \textbf{2014}, \emph{251}, 2545--2550\relax
\mciteBstWouldAddEndPuncttrue
\mciteSetBstMidEndSepPunct{\mcitedefaultmidpunct}
{\mcitedefaultendpunct}{\mcitedefaultseppunct}\relax
\EndOfBibitem
\bibitem[Engels et~al.(2014)Engels, Terr{\'e}s, Epping, Khodkov, Watanabe,
  Taniguchi, Beschoten, and Stampfer]{engels2014}
Engels,~S.; Terr{\'e}s,~B.; Epping,~A.; Khodkov,~T.; Watanabe,~K.;
  Taniguchi,~T.; Beschoten,~B.; Stampfer,~C. \emph{Physical Review Letters}
  \textbf{2014}, \emph{113}, 126801\relax
\mciteBstWouldAddEndPuncttrue
\mciteSetBstMidEndSepPunct{\mcitedefaultmidpunct}
{\mcitedefaultendpunct}{\mcitedefaultseppunct}\relax
\EndOfBibitem
\bibitem[Neumann et~al.(2015)Neumann, Reichardt, Dr{\"o}geler, Terr{\'e}s,
  Watanabe, Taniguchi, Beschoten, Rotkin, and Stampfer]{neumann2015}
Neumann,~C.; Reichardt,~S.; Dr{\"o}geler,~M.; Terr{\'e}s,~B.; Watanabe,~K.;
  Taniguchi,~T.; Beschoten,~B.; Rotkin,~S.~V.; Stampfer,~C. \emph{Nano Letters}
  \textbf{2015}, \emph{15}, 1547--1552\relax
\mciteBstWouldAddEndPuncttrue
\mciteSetBstMidEndSepPunct{\mcitedefaultmidpunct}
{\mcitedefaultendpunct}{\mcitedefaultseppunct}\relax
\EndOfBibitem
\bibitem[Kretinin et~al.(2014)Kretinin, Cao, Tu, Yu, Jalil, Novoselov, Haigh,
  Gholinia, Mishchenko, Lozada, Georgiou, Woods, Withers, Blake, Eda, Wirsig,
  Hucho, Watanabe, Taniguchi, Geim, and Gorbachev]{kretinin2014}
Kretinin,~A. et~al.  \emph{Nano Letters} \textbf{2014}, \emph{14},
  3270--3276\relax
\mciteBstWouldAddEndPuncttrue
\mciteSetBstMidEndSepPunct{\mcitedefaultmidpunct}
{\mcitedefaultendpunct}{\mcitedefaultseppunct}\relax
\EndOfBibitem
\bibitem[Dean et~al.(2010)Dean, Young, Meric, Lee, Wang, Sorgenfrei, Watanabe,
  Taniguchi, Kim, Shepard, and Hone]{dean2010}
Dean,~C.; Young,~A.; Meric,~I.; Lee,~C.; Wang,~L.; Sorgenfrei,~S.;
  Watanabe,~K.; Taniguchi,~T.; Kim,~P.; Shepard,~K.; Hone,~J. \emph{Nature
  Nanotechnology} \textbf{2010}, \emph{5}, 722--726\relax
\mciteBstWouldAddEndPuncttrue
\mciteSetBstMidEndSepPunct{\mcitedefaultmidpunct}
{\mcitedefaultendpunct}{\mcitedefaultseppunct}\relax
\EndOfBibitem
\bibitem[Britnell et~al.(2012)Britnell, Gorbachev, Jalil, Belle, Schedin,
  Mishchenko, Georgiou, Katsnelson, Eaves, Morozov, Peres, Leist, Geim,
  Novoselov, and Ponomarenko]{britnell2012}
Britnell,~L.; Gorbachev,~R.; Jalil,~R.; Belle,~B.; Schedin,~F.; Mishchenko,~A.;
  Georgiou,~T.; Katsnelson,~M.; Eaves,~L.; Morozov,~S.; Peres,~N.; Leist,~J.;
  Geim,~A.; Novoselov,~K.; Ponomarenko,~L. \emph{Science} \textbf{2012},
  \emph{335}, 947--950\relax
\mciteBstWouldAddEndPuncttrue
\mciteSetBstMidEndSepPunct{\mcitedefaultmidpunct}
{\mcitedefaultendpunct}{\mcitedefaultseppunct}\relax
\EndOfBibitem
\bibitem[Couto et~al.(2014)Couto, Costanzo, Engels, Ki, Watanabe, Taniguchi,
  Stampfer, Guinea, and Morpurgo]{couto2014}
Couto,~N.~J.; Costanzo,~D.; Engels,~S.; Ki,~D.-K.; Watanabe,~K.; Taniguchi,~T.;
  Stampfer,~C.; Guinea,~F.; Morpurgo,~A.~F. \emph{Physical Review X}
  \textbf{2014}, \emph{4}, 041019\relax
\mciteBstWouldAddEndPuncttrue
\mciteSetBstMidEndSepPunct{\mcitedefaultmidpunct}
{\mcitedefaultendpunct}{\mcitedefaultseppunct}\relax
\EndOfBibitem
\bibitem[Dr{\"o}geler et~al.(2014)Dr{\"o}geler, Volmer, Wolter, Terr{\'e}s,
  Watanabe, Taniguchi, G{\"u}ntherodt, Stampfer, and Beschoten]{drogeler2014}
Dr{\"o}geler,~M.; Volmer,~F.; Wolter,~M.; Terr{\'e}s,~B.; Watanabe,~K.;
  Taniguchi,~T.; G{\"u}ntherodt,~G.; Stampfer,~C.; Beschoten,~B. \emph{Nano
  Letters} \textbf{2014}, \emph{14}, 6050--6055\relax
\mciteBstWouldAddEndPuncttrue
\mciteSetBstMidEndSepPunct{\mcitedefaultmidpunct}
{\mcitedefaultendpunct}{\mcitedefaultseppunct}\relax
\EndOfBibitem
\bibitem[Ju et~al.(2014)Ju, Velasco~Jr, Huang, Kahn, Nosiglia, Tsai, Yang,
  Taniguchi, Watanabe, Zhang, Zhang, Crommie, Zettl, and Wang]{ju2014}
Ju,~L.; Velasco~Jr,~J.; Huang,~E.; Kahn,~S.; Nosiglia,~C.; Tsai,~H.-Z.;
  Yang,~W.; Taniguchi,~T.; Watanabe,~K.; Zhang,~Y.; Zhang,~G.; Crommie,~G.;
  Zettl,~A.; Wang,~F. \emph{Nature Nanotechnology} \textbf{2014}, \emph{9},
  348--352\relax
\mciteBstWouldAddEndPuncttrue
\mciteSetBstMidEndSepPunct{\mcitedefaultmidpunct}
{\mcitedefaultendpunct}{\mcitedefaultseppunct}\relax
\EndOfBibitem
\bibitem[Kim et~al.(2013)Kim, Bae, Seo, Kim, Kim, Lee, Ahn, Lee, Chun, and
  Park]{kim2013b}
Kim,~Y.~D.; Bae,~M.-H.; Seo,~J.-T.; Kim,~Y.~S.; Kim,~H.; Lee,~J.~H.;
  Ahn,~J.~R.; Lee,~S.~W.; Chun,~S.-H.; Park,~Y.~D. \emph{ACS Nano}
  \textbf{2013}, \emph{7}, 5850--5857\relax
\mciteBstWouldAddEndPuncttrue
\mciteSetBstMidEndSepPunct{\mcitedefaultmidpunct}
{\mcitedefaultendpunct}{\mcitedefaultseppunct}\relax
\EndOfBibitem
\bibitem[Tiberj et~al.(2013)Tiberj, Rubio-Roy, Paillet, Huntzinger, Landois,
  Mikolasek, Contreras, Sauvajol, Dujardin, and Zahab]{tiberj2013}
Tiberj,~A.; Rubio-Roy,~M.; Paillet,~M.; Huntzinger,~J.-R.; Landois,~P.;
  Mikolasek,~M.; Contreras,~S.; Sauvajol,~J.-L.; Dujardin,~E.; Zahab,~A.-A.
  \emph{Scientific Reports} \textbf{2013}, \emph{3}, 2355\relax
\mciteBstWouldAddEndPuncttrue
\mciteSetBstMidEndSepPunct{\mcitedefaultmidpunct}
{\mcitedefaultendpunct}{\mcitedefaultseppunct}\relax
\EndOfBibitem
\bibitem[Banszerus et~al.(2015)Banszerus, Schmitz, Engels, Dauber, Oellers,
  Haupt, Watanabe, Taniguchi, Beschoten, and Stampfer]{banszerus2015}
Banszerus,~L.; Schmitz,~M.; Engels,~S.; Dauber,~J.; Oellers,~M.; Haupt,~F.;
  Watanabe,~K.; Taniguchi,~T.; Beschoten,~B.; Stampfer,~C. \emph{Science
  Advances} \textbf{2015}, \emph{1}, e1500222\relax
\mciteBstWouldAddEndPuncttrue
\mciteSetBstMidEndSepPunct{\mcitedefaultmidpunct}
{\mcitedefaultendpunct}{\mcitedefaultseppunct}\relax
\EndOfBibitem
\bibitem[Neumann et~al.(2015)Neumann, Reichardt, Venezuela, Dr{\"o}geler,
  Banszerus, Schmitz, Watanabe, Taniguchi, Mauri, Beschoten, Rotkin, and
  Stampfer]{neumann2015b}
Neumann,~C.; Reichardt,~S.; Venezuela,~P.; Dr{\"o}geler,~M.; Banszerus,~L.;
  Schmitz,~M.; Watanabe,~K.; Taniguchi,~T.; Mauri,~F.; Beschoten,~B.;
  Rotkin,~S.~V.; Stampfer,~C. \emph{Nature Communications
	} \textbf{2015}, \emph{6}, 8429\relax
\mciteBstWouldAddEndPuncttrue
\mciteSetBstMidEndSepPunct{\mcitedefaultmidpunct}
{\mcitedefaultendpunct}{\mcitedefaultseppunct}\relax
\EndOfBibitem
\bibitem[Attaccalite et~al.(2011)Attaccalite, Bockstedte, Marini, Rubio, and
  Wirtz]{attaccalite2011}
Attaccalite,~C.; Bockstedte,~M.; Marini,~A.; Rubio,~A.; Wirtz,~L.
  \emph{Physical Review B} \textbf{2011}, \emph{83}, 144115\relax
\mciteBstWouldAddEndPuncttrue
\mciteSetBstMidEndSepPunct{\mcitedefaultmidpunct}
{\mcitedefaultendpunct}{\mcitedefaultseppunct}\relax
\EndOfBibitem
\bibitem[Novoselov et~al.(2004)Novoselov, Geim, Morozov, Jiang, Zhang, Dubonos,
  Grigorieva, and Firsov]{novoselov2004}
Novoselov,~K.~S.; Geim,~A.~K.; Morozov,~S.; Jiang,~D.; Zhang,~Y.; Dubonos,~S.;
  Grigorieva,~I.; Firsov,~A. \emph{Science} \textbf{2004}, \emph{306},
  666--669\relax
\mciteBstWouldAddEndPuncttrue
\mciteSetBstMidEndSepPunct{\mcitedefaultmidpunct}
{\mcitedefaultendpunct}{\mcitedefaultseppunct}\relax
\EndOfBibitem
\bibitem[Williams et~al.(2007)Williams, DiCarlo, and Marcus]{williams2007}
Williams,~J.; DiCarlo,~L.; Marcus,~C. \emph{Science} \textbf{2007}, \emph{317},
  638--641\relax
\mciteBstWouldAddEndPuncttrue
\mciteSetBstMidEndSepPunct{\mcitedefaultmidpunct}
{\mcitedefaultendpunct}{\mcitedefaultseppunct}\relax
\EndOfBibitem
\bibitem[Rickhaus et~al.(2013)Rickhaus, Maurand, Liu, Weiss, Richter, and
  Sch{\"o}nenberger]{rickhaus2013}
Rickhaus,~P.; Maurand,~R.; Liu,~M.-H.; Weiss,~M.; Richter,~K.;
  Sch{\"o}nenberger,~C. \emph{Nature Communications} \textbf{2013}, \emph{4},
  2342\relax
\mciteBstWouldAddEndPuncttrue
\mciteSetBstMidEndSepPunct{\mcitedefaultmidpunct}
{\mcitedefaultendpunct}{\mcitedefaultseppunct}\relax
\EndOfBibitem
\bibitem[Young and Kim(2009)Young, and Kim]{young2009}
Young,~A.~F.; Kim,~P. \emph{Nature Physics} \textbf{2009}, \emph{5},
  222--226\relax
\mciteBstWouldAddEndPuncttrue
\mciteSetBstMidEndSepPunct{\mcitedefaultmidpunct}
{\mcitedefaultendpunct}{\mcitedefaultseppunct}\relax
\EndOfBibitem
\bibitem[Taychatanapat et~al.(2015)Taychatanapat, Tan, Yeo, Watanabe,
  Taniguchi, and {\"O}zyilmaz]{taychatanapat2015}
Taychatanapat,~T.; Tan,~J.~Y.; Yeo,~Y.; Watanabe,~K.; Taniguchi,~T.;
  {\"O}zyilmaz,~B. \emph{Nature Communications} \textbf{2015}, \emph{6},
  6093\relax
\mciteBstWouldAddEndPuncttrue
\mciteSetBstMidEndSepPunct{\mcitedefaultmidpunct}
{\mcitedefaultendpunct}{\mcitedefaultseppunct}\relax
\EndOfBibitem
\bibitem[Cheianov et~al.(2007)Cheianov, Fal'ko, and Altshuler]{cheianov2007}
Cheianov,~V.~V.; Fal'ko,~V.; Altshuler,~B. \emph{Science} \textbf{2007},
  \emph{315}, 1252--1255\relax
\mciteBstWouldAddEndPuncttrue
\mciteSetBstMidEndSepPunct{\mcitedefaultmidpunct}
{\mcitedefaultendpunct}{\mcitedefaultseppunct}\relax
\EndOfBibitem
\bibitem[Beenakker(2008)]{beenakker2008}
Beenakker,~C. \emph{Reviews of Modern Physics} \textbf{2008}, \emph{80},
  1337\relax
\mciteBstWouldAddEndPuncttrue
\mciteSetBstMidEndSepPunct{\mcitedefaultmidpunct}
{\mcitedefaultendpunct}{\mcitedefaultseppunct}\relax
\EndOfBibitem
\end{mcitethebibliography}
\end{document}